\documentclass{article}
\usepackage{amsmath,graphicx}
\usepackage[english]{babel}
\usepackage{hyperref}
\usepackage{amssymb}
\usepackage{booktabs}
\usepackage{support-caption}
\usepackage{subcaption}
\usepackage[preprint]{spconf}
\toappear{\parbox{\linewidth}{\textbf{This work has been submitted to the IEEE for possible publication. Copyright may be transferred without notice, after which this version may no longer be accessible.}}}
\usepackage{enumitem}
\setlist{nosep, leftmargin=14pt}

\usepackage{mwe} 

\title{Novel structural-scale uncertainty measures and error retention curves: application to multiple sclerosis}

\name{\parbox{\linewidth}{Nataliia Molchanova$^{1,2,3}$ \quad Vatsal Raina$^{3,4}$ \quad Andrey Malinin$^{5}$ \quad Francesco La Rosa$^{6}$ \\ \textit{Henning Muller}$^{3}$ \quad \textit{Mark Gales}$^{4}$ \quad \textit{Cristina Granziera}$^{7}$ \quad \textit{Mara Graziani}$^{3,8}$ \quad \textit{Meritxell Bach Cuadra}$^{1,2}$}}

\address{
$^{1}$ Lausanne University Hospital, Switzerland, $^{2}$ University of Lausanne, Switzerland \\
$^{3}$ University of Applied Sciences of Western Switzerland, Switzerland, 
$^{4}$ University of Cambridge, UK, \\
$^{5}$ Shifts Project, Finland, $^{6}$ Icahn School of Medicine at Mount Sinai, USA, \\
$^{7}$ University Hospital Basel, Switzerland, $^{8}$ IBM Research Europe, Switzerland.
}
\begin{document}
%
\maketitle
\begin{abstract}
This paper focuses on the uncertainty estimation for white matter lesions (WML) segmentation in magnetic resonance imaging (MRI). On one side, voxel-scale segmentation errors cause the erroneous delineation of the lesions; on the other side, lesion-scale detection errors lead to wrong lesion counts. Both of these factors are clinically relevant for the assessment of multiple sclerosis patients. This work aims to compare the ability of different voxel- and lesion-scale uncertainty measures to capture errors related to segmentation and lesion detection, respectively. Our main contributions are (i) proposing new measures of lesion-scale uncertainty that do not utilise voxel-scale uncertainties; (ii) extending an error retention curves analysis framework for evaluation of lesion-scale uncertainty measures. Our results obtained on the multi-center testing set of 58 patients demonstrate that the proposed lesion-scale measure achieves the best performance among the analysed measures. All code implementations are provided at \url{https://github.com/NataliiaMolch/MS_WML_uncs}.
\end{abstract}
%
\section{Introduction}
\label{sec:intro}

Magnetic resonance imaging (MRI) is a key imaging tool used for multiple sclerosis (MS) diagnosis and prognosis~\cite{hemond_magnetic_2018}. Brain white matter lesions (WML), visible in MRI, are included in the McDonald MS diagnostic criteria and used for the assessment of disease progression~\cite{mac, bendfeldt_association_2009, fisniku_disability_2008}. Both dissemination in space and time of WML in MRI are needed in clinical assessment of MS patients~\cite{mac}.

Given the variability of lesion sizes, locations and per-patient counts, manual WML annotation is a skill-demanding time-consuming task, yet prone to human errors. Automatic segmentation methods can speed up the annotation process and reduce an annotator bias. Deep learning models constitute the state of the art for MS lesion segmentation~\cite{mssegreview}. Uncertainty quantification has been previously explored for this task to provide clinicians with information about the reliability and confidence of predictions~\cite{TAlber, MCKINLEY2020102104, Dojat, DojatISMRM, DojatMICCAI}. Various uncertainty measures of voxel-scale predictions were explored, such as entropy, variance, predicted confidence and mutual information~\cite{TAlber, DojatISMRM, DojatMICCAI}. The same studies suggested evaluating uncertainty measures on a structural scale, \textit{i.e.} for each predicted lesion, by aggregating voxel-scale uncertainties. Some attempts were made to quantify the informativeness of uncertainty measures on voxel and lesion scales~\cite{TAlber, DojatMICCAI}. For instance, \cite{TAlber} looked into uncertainty-based prediction filtering as a means to correlate uncertainty and false positive errors, and \cite{DojatMICCAI} used accuracy-confidence curves. However, these approaches do not explicitly assess the correspondence between the measures and model errors in segmentation or lesion detection, while such an analysis is essential to decide what uncertainty maps should be presented to clinicians.

\begin{figure}[htb]
  \centering
  \vspace{-0.1cm}
  \centerline{\includegraphics[width=8cm]{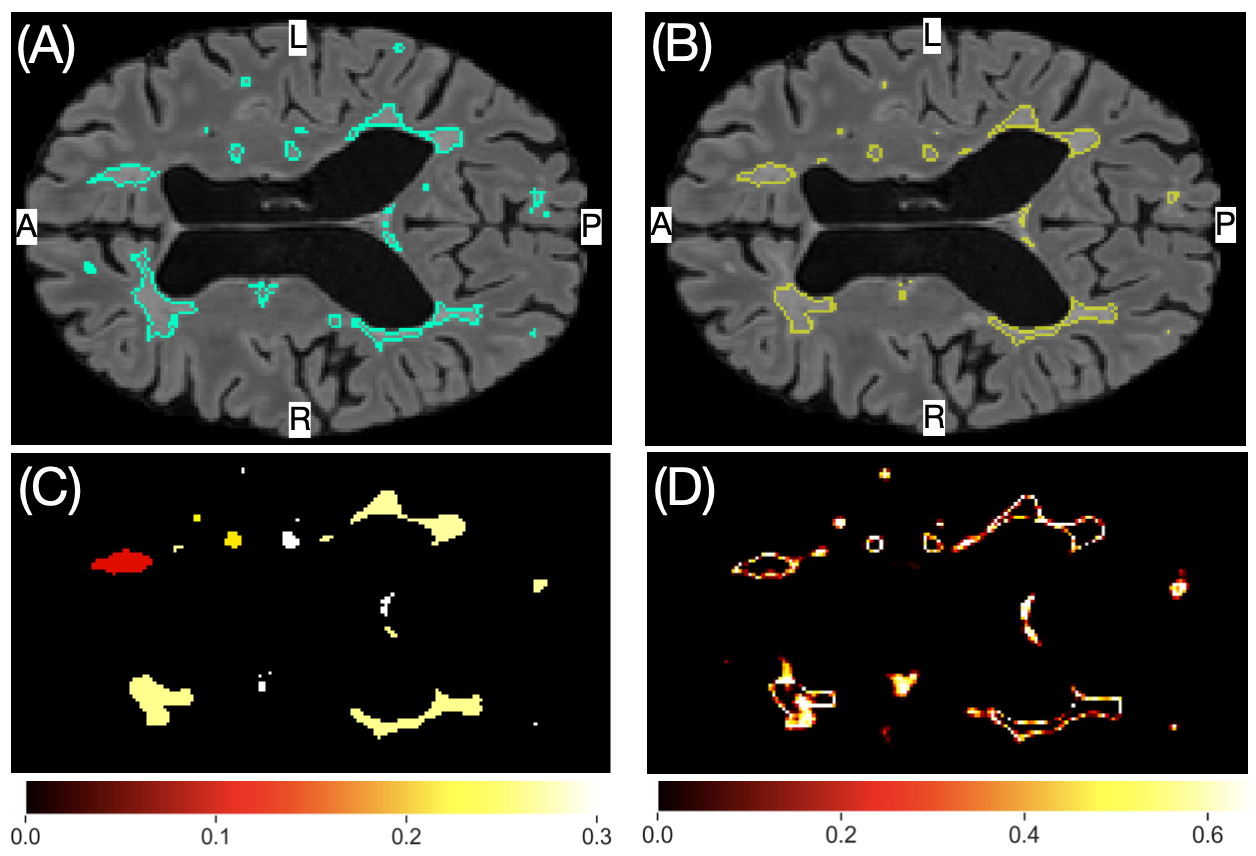}}
\caption{An example of WML and uncertainty maps. \textbf{(A)} and \textbf{(B)} are ground truth (blue) and predicted (yellow) WML borders overlayed on a FLAIR scan, \textbf{(C)} and \textbf{(D)} lesion-scale (proposed DDU$^{true}$) and voxel-scale (EoE) uncertainty maps.}
\label{fig:ex}
\end{figure}

In this work, we promote the use of error retention curves (RC)~\cite{malinin2019uncertainty} for assessing voxel- and lesion-scale uncertainty measures. We first compare the ability of six voxel-scale measures to capture errors related to the overall segmentation. Our first contribution is novel lesion-scale uncertainty measures that, in contrast to previously proposed ones~\cite{TAlber, DojatISMRM}, are not based on voxel-scale uncertainties, but use structural information directly. The second contribution is the extension of the RC definition to the structural scale to quantify the correspondence between lesion-scale uncertainty measures and lesion detection errors. With this framework, we compare the proposed lesion-scale uncertainty measures to previously proposed ones.
In particular, our results show a superior performance of the proposed lesion-scale measure in comparison to the measures based on the voxel-scale uncertainties.
%
\section{Methodology}
\label{sec:methods}
\subsection{Voxel-scale uncertainty}
\label{sec:voxuncs}
In this work, we use deep ensembles for uncertainty quantification, which have shown to provide better uncertainty estimates for the WML segmentation task than a more common Monte Carlo dropout~\cite{DojatISMRM, shifts20}. In a classification tasks, several measures of predictive uncertainty can be approximated using predicted probability distributions over the classes from each of the ensemble members. Such uncertainty measures are categorised as data, knowledge and total uncertainty \cite{malinin2019uncertainty}. Data uncertainty measures the inherent noise within the source data distribution. Knowledge uncertainty captures the disagreement in the outputs of each model in the ensemble. This reflects the lack of knowledge by the model in certain regions of the input space. Finally, total uncertainty considers the contributions of both data and knowledge uncertainty.

\begin{table}[b!]
\fontsize{8}{8}\selectfont
\centering
\begin{small}
    \begin{tabular}{l|l}
    \toprule
\multicolumn{2}{c}{Total uncertainty} \\ \midrule
EoE & $-\sum_y \frac{1}{K}\sum_{k=1}^K P_{k}(\mathbf{y}) \log\left[\frac{1}{K}\sum_{k=1}^K P_{k}(\mathbf{y})\right]$ \\
NC & $-\operatorname*{argmax}_y \frac{1}{K}\sum_{k=1}^K P_k(\mathbf{y})$ \\
\midrule
\multicolumn{2}{c}{Data uncertainty} \\ \midrule
ExE & $ -\frac{1}{K}\sum_{k=1}^K \sum_y P_{k}(\mathbf{y})\log P_{k}(\mathbf{y}) $ \\
\midrule
\multicolumn{2}{c}{Knowledge uncertainty} \\ \midrule
MI & EoE $-$ ExE \\
EPKL & $-\frac{1}{K^2}\sum_y\left[ \sum_{k=1}^K P_{k}(\mathbf{y}) \sum_{k=1}^K \log P_{k}(\mathbf{y})\right]$ $-$ ExE \\
RMI & EPKL - MI \\
  \bottomrule
    \end{tabular}
    \end{small}
\caption{Definitions of voxel-scale uncertainty measures\cite{malinin2020Struct}. Notations: $P_k(\mathbf{y}) \equiv P(\mathbf{y}|\mathbf{x}, \mathbf{\theta}_{k})$ - a predictive posterior of the $k^{th}$ model in the ensemble of size $K$, $\mathbf{y}$ - vector of model's outputs, $\mathbf{x}$ - vector of inputs, $\mathbf{\theta}_{k}$ - weights of the $k^{th}$ model sampled from a posterior $q(\mathbf{\theta})$.}
\label{tab:vox_unc}
\end{table}

\pagebreak
WML segmentation can be seen as distinct binary classification tasks for each voxel. Thus, voxel-scale uncertainty maps can be calculated using standard measures of data, knowledge and total uncertainty. We estimate the following uncertainty measures (definitions in Table~\ref{tab:vox_unc}): mutual information (MI), expected pair-wise KL divergence (EPKL) and reverse mutual information (RMI) for knowledge uncertainty; expected entropy (ExE) for data uncertainty; entropy of expected (EoE) and negated confidence (NC) for total uncertainty~\cite{malinin2020Struct}. The entropy-based measures, MI and NC have previously been investigated in MS by~\cite{TAlber, DojatISMRM, DojatMICCAI} while EPKL and RMI are investigated for the first time. 
\vspace{-0.2cm}
\subsection{Lesion-scale uncertainty}
Previously considered lesion-scale uncertainty measures aggregate information from a voxel-scale uncertainty map $U \in \mathbb{R}^{H \times W \times D}$, by computing either mean~\cite{DojatISMRM} or log-sum~\cite{TAlber} of $U$ values within a predicted lesion region $\Omega$, as $\frac{1}{|\Omega|}\sum_{i \in \Omega}U_i$ and $\sum_{i \in \Omega}logU_i$, where $U_i$ is an uncertainty of the $i^{th}$ voxel of a lesion region $\Omega$. Lesion regions $\Omega$ are determined as connected components on a predicted binary lesion map. The voxel-scale uncertainty map $U$ can be obtained using different measures, we explore all the measures defined in Table~\ref{tab:vox_unc}.

Previous investigations showed that the highest uncertainty regions are usually concentrated on the borders of lesions~\cite{MCKINLEY2020102104, Dojat, shifts20} (see Figure \ref{fig:ex} D). Thus, voxel-scale uncertainties are more likely to be informative for lesion delineation, and may not be so relevant for lesion detection.
From this rationale, we propose two novel structural-scale measures that do not use the voxel-scale uncertainties, but rather look at the disagreement between different models in an ensemble about structural predictions. Let $\Omega$ be a lesion region predicted by an ensemble, $\Omega_k$ - a region of the same lesion predicted by the $k^{th}$ ensemble member ($k=1,2,...,K$), \textit{i.e.} a connected component on the binary mask predicted by the $k^{th}$ model having the maximum intersection over union (IoU) with $\Omega$. The proposed measure of the disagreement between the models in the ensemble on the detected area, further referred to as detection disagreement uncertainty (DDU). DDU is defined as one minus agreement, where agreement is an average across ensemble members IoU between the lesion regions predicted by the ensemble and individual ensemble members:
\begin{equation}
\text{DDU} = 1-\frac{1}{K} \sum_{k=1}^{K} \text{IoU}(\Omega,\Omega_{k})
\label{eq:ddu}
\end{equation}

We tested two ways of obtaining binary segmentation maps from ensemble members, which yield different definitions of lesion regions $\Omega_k$: (i) applying the probability threshold tuned for the ensemble to individual model predictions, (ii) tuning the probability thresholds separately for each model in the ensemble. Resulting measures will be referred as DDU for (i) and DDU$^{true}$ for (ii).
\subsection{Error retention curves for uncertainty evaluation}
Ideally, a high uncertainty score should highlight the predictions that are the most likely to be wrong. Hence, we expect a good quality uncertainty measure to reflect the increased likelihood of an erroneous prediction and thus correlate with model mistakes. The error RC~\cite{malinin2019uncertainty, shifts20, brats} assesses the degree of this relation by only looking at the ranking of uncertainties. 
On the voxel scale, RCs analyse the correspondence between a measure of uncertainty and the model errors in segmentation (\textit{e.g.} Dice Similarity Coefficient, DSC). The RC for a single scan is built by sequentially replacing a fraction $\tau$ of the most uncertain voxel predictions within the brain mask with the ground truth and re-computing the DSC. The area under the DSC retention curve (DSC-RC), further referred to as DSC-AUC, is a measure of correspondence between an uncertainty measure and segmentation errors that only takes into account the ranking of uncertainties in a particular scan. To summarise the performance on the whole dataset, an average across patients DSC-RC and area under it, \textit{i.e.} $\widehat{\text{DSC-AUC}}$, are computed. It is also possible to estimate lower and upper bounds of performance by building \emph{random} and \emph{ideal} RCs. For a random RC, we assign random uncertainty values to each voxel of predictions. For the ideal one, a zero uncertainty is assigned to true positive and negative (TP and TN) voxels and while false positive and negative (FP and FN) voxels have an uncertainty of 1. To build the RCs, we use $\tau=2.5\cdot10^{-3}$.

RCs are still unstudied at the scale of structural predictions, \textit{i.e.} a lesion scale in our case. 
We propose to build structural-scale RCs to see the how lesion-scale uncertainty measures are associated with the lesion detection errors. We use a lesion F1-score~\cite{mssegchallenge} to measure the detection quality. It is defined as $\frac{2TP}{2TP + FN + FP}$, where $TP$ is a count of predicted lesions that have maximum across the ground truth lesions IoU greater than $\gamma$; inversely FP lesions have a maximum IoU lower than $\gamma$; $FN$ is the count of ground truth lesions not overlapping with predictions. Usually a few voxels overlap is considered enough for lesion detection~\cite{TAlber, CARASS201777, LAROSA2020102335}, however we promote a more rigorous criteria and use $\gamma=0.25$.

Novel analysis of uncertainty with lesion-scale F1-RC poses new challenges. First, compared to voxel-scale predictions, the amount of predicted lesions is much smaller and varies from patient to patient. Therefore, we sequentially reject \emph{each} lesion in a scan based on its uncertainty ranking. Secondly, it is only possible to compute the uncertainties of predicted lesions, \textit{i.e.} TP and FP, as FN lesion regions are defined from the ground truth and TN lesions do not exist. Thus, we suggest the following scheme of sequential lesion rejection: TP lesions remain TP, and FP lesions are just excluded. Note that FN lesions are not replaced in the F1-RC, and the $FN$ count remains fixed in the F1-score formula. 
Thirdly, lesion F1-RC are computed for each patient in different sets of retention fractions due to varying lesion count across patients. 
To average subject-wise F1-RCs, we use a piece-wise linear interpolation to interpolate the obtained curves to a similar set of retention fractions. It allows averaging the F1-RCs at each node to obtain an average across-dataset F1-RC. The lesion $\widehat{\text{F1-AUC}}$ is computed by averaging the F1-AUC of individual patients after interpolation. $\widehat{\text{F1-AUC}}$ quantifies the correspondence between a lesion-scale uncertainty measure and lesion detection errors in a particular dataset. In analogy to the DSC-RC, the ideal and random retention curves are built.
\subsection{Dataset}
We used a publicly available dataset provided by the Shifts project~\cite{shifts20}, designed for the study of uncertainty estimation across shifted domains (different locations, scanners, MS stages, \textit{etc.}). It is composed of fluid-attenuated inversion recovery (FLAIR)~\cite{flair} scans, which underwent denoising, skull stripping, bias field correction and interpolation to $1 mm$ iso-voxel space, and their manual WML annotations used as the ground truth. Training and validation datasets contain data from four different centers with 33 and 7 scans, respectively. For testing we used a total of 58 subjects, which comprise both in-domain data sampled from the same distribution as the training one (33 cases) and out-of-domain data collected in a different center and scanner (25 cases). The combination of in- and out-of-domain data is beneficial for RC construction, as it helps to embrace a greater amount of model errors (errors made on both in-domain and out-of-domain data), thus a more reliable evaluation of the uncertainty measures is obtained.
\subsection{Models}
We use a state-of-the-art WML segmentation method based on 3D U-net architecture~\cite{LAROSA2020102335}, which was previously used in uncertainty studies for the particular task~\cite{TAlber, DojatISMRM, shifts20, DojatMICCAI}. Our final model is an ensemble of five U-nets ($K=5$), each trained with different random seed initialisation. Training of each model is performed for a maximum of 300 epochs with the best model selection based on validation performance. A combination of Dice and focal losses is used as an objective function optimised using an Adam algorithm with a constant learning rate of $10^{-5}$. The architecture takes as an input $96\times96\times96$ sub-volumes of FLAIR, which are aggregated using Gaussian-weighted averaging at inference time. Binary WML segmentation masks are then obtained by thresholding the average across ensemble members probability maps at a value of 0.3 chosen on a validation set by optimising the DSC. For DDU$^{true}$ measure (eq. \ref{eq:ddu}) computation, binary segmentation maps are obtained from each ensemble member using the thresholds of $0.35, 0.25, 0.25, 0.25, 0.1$ chosen separately for each ensemble member based on DSC.
\section{Results}
\begin{figure}[htb]
\begin{minipage}[b]{1.0\linewidth}
  \centering
  \centerline{(a) DSC-RC}\medskip
  \centerline{\includegraphics[width=7cm]{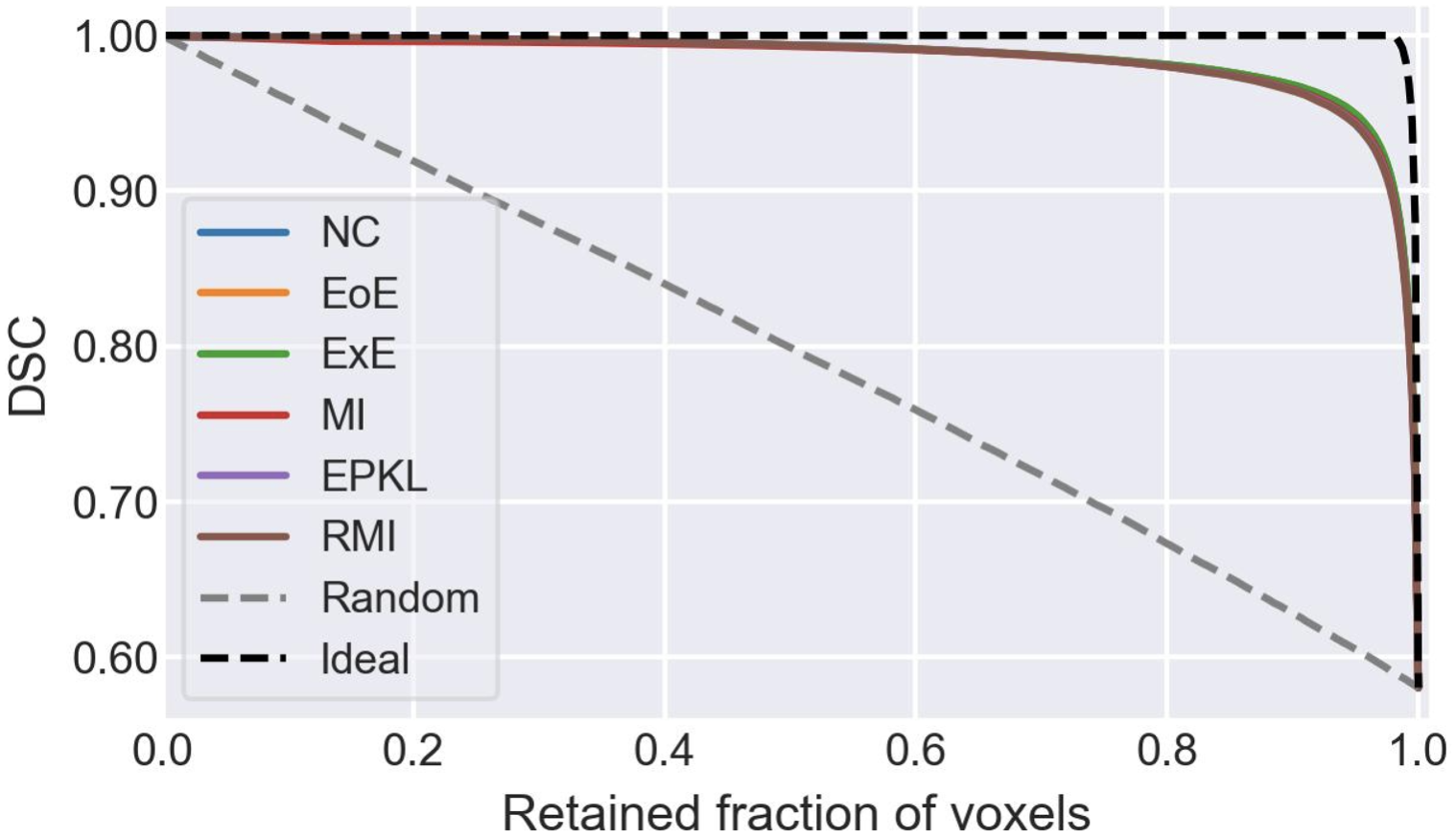}}
\end{minipage}
\begin{minipage}[b]{1.0\linewidth}
  \centering
  \vspace{0.2cm}
  \centerline{(b) Lesion F1-RC}\medskip
  \centerline{\includegraphics[width=7cm]{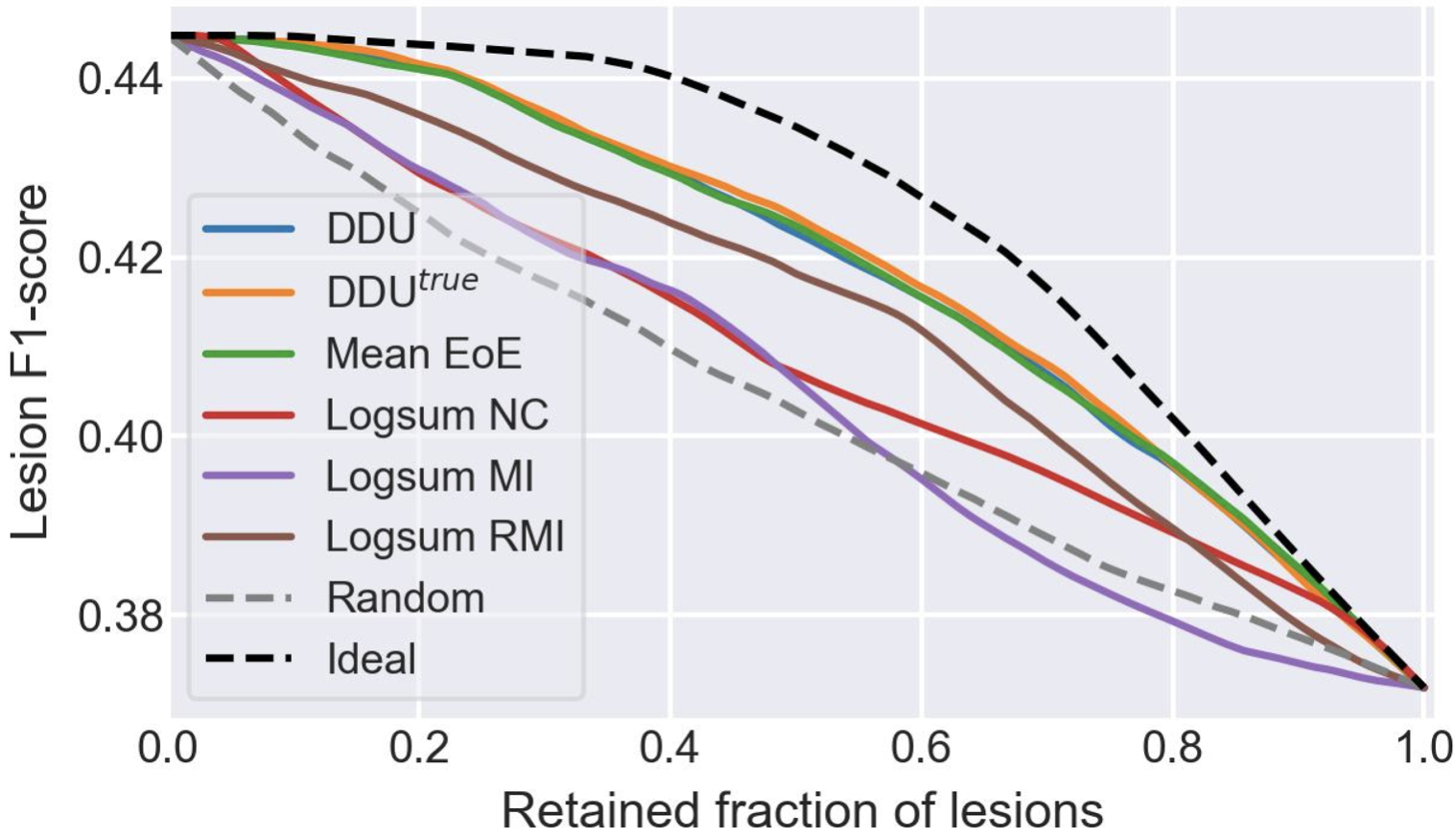}}
\end{minipage}
\caption{Average across the test set DSC (a) and lesion F1 (b) retention curves for evaluation of uncertainty measures on voxel and lesion scales, respectively. Lesion F1-RC are made for the three best (Mean EoE, DDU, Mean NC) and the three worst (Logsum RMI, NC, MI) measures in terms of $\widehat{\text{F1-AUC}}$.}
\label{fig:res}
\end{figure}

The resulting average RCs are shown in Figure~\ref{fig:res} and the corresponding AUCs are given in Table~\ref{tab:aac}. At the voxel-scale, ExE and EoE measures show the highest $\widehat{\text{DSC-AUC}}$ at the level of $0.9853_{\pm 0.002}$, meaning the best ability to capture errors in overall segmentation. Nevertheless, all uncertainty measures show comparable results that are closer to ideal performance than random. On the lesion scale, the performance of log-sum-based measures (the only ones highly dependent on lesion size) is the poorest in terms of $\widehat{\text{F1-AUC}}$. The best performance is given by the proposed DDU$^{true}$ with $\widehat{\text{F1-AUC}}$ of $0.4265_{\pm 0.0294}$, meaning the best ability to capture lesion detection errors. The performance of Mean EoE, DDU and Mean NC measures is close to DDU$^{true}$, but it is statistically significantly different at significance level of 0.05 according to one-sided paired Wilcoxon tests that show p-values of 0.007, 0.013 and 0.003, respectively.
\section{Discussion}
In this work, we propose novel lesion-scale uncertainty measures that, in comparison to previously proposed ones, do not utilise voxel-scale uncertainties and are rather defined through the disagreement between ensemble members in structural predictions. Differently from previous works exploring uncertainty in MS, we used error retention curves to evaluate a range of uncertainty measures in their ability to capture model errors on both voxel and lesion scales. The comparison between uncertainty measures was performed in terms of their ability to capture overall segmentation or lesion detection errors via DSC-RC and lesion F1-RC, respectively.  

We observe that a comparatively good performance of an uncertainty measure on the voxel scale does not mean that it will be useful for pinpointing lesion detection errors. For example, while ExE (data uncertainty) and EoE (total uncertainty) measures have a similar and superior performance in terms of $\widehat{\text{DSC-AUC}}$, the Mean ExE is the least informative for the lesion detection across mean-based lesion-scale measures. Additionally, we notice that a previously proposed log-sum aggregation strategy for lesion-scale measures construction yields one of the lowest $\widehat{\text{F1-AUC}}s$. In contrast to lesion-scale measures based on voxel uncertainties, the proposed DDU$^{true}$ shows significantly better performance in capturing lesion detection errors, measured by $\widehat{\text{F1-AUC}}$.
The proposed framework can be a step forward in integrating AI in a clinical practice considering a semi-automatic scenario where clinicians assess automated WML predictions and make necessary corrections. Although we identified uncertainty measures informative in terms of model errors of different kind, it is yet important to verify in practice if introducing uncertainty maps to clinicians can speed up or simplify the correction process of predicted WML masks.

\begin{table}[hb]
    \begin{subtable}[t]{.49\linewidth}
      \centering
      \small
      \caption{$\widehat{\text{DSC-AUC}} * 100 (\uparrow)$}
 \resizebox{.9\textwidth}{!}{%
        \begin{tabular}{ll}
    \toprule
    Ideal & $99.89_{\pm 0.01}$ \\
    ExE & $98.53_{\pm 0.20}$ \\
    EoE & $98.53_{\pm 0.20}$ \\
    NC & $98.50_{\pm 0.20}$ \\
    EPKL & $98.44_{\pm 0.21}$ \\
    RMI & $98.41_{\pm 0.21}$ \\
    MI & $98.37_{\pm 0.21}$ \\
    Random & $79.66_{\pm 1.46}$ \\
    \bottomrule
    \end{tabular}}
    \end{subtable}%
    \begin{subtable}[t]{.49\linewidth}
      \centering
        \caption{Lesion $\widehat{\text{F1-AUC}} * 100 (\uparrow)$}
        \resizebox{.9\textwidth}{!}{%
        \begin{tabular}{ll}
            \toprule
            Ideal & $43.22_{\pm 2.99}$ \\
            DDU$^{true}$ (\textbf{our}) & $42.65_{\pm 2.94}$ \\
            Mean EoE & $42.60_{\pm 2.94}$ \\
            DDU (\textbf{our}) & $42.59_{\pm 2.94}$ \\
            Mean NC & $42.59_{\pm 2.94}$ \\
            Mean MI & $42.43_{\pm 2.92}$ \\
            Mean EPKL & $42.42_{\pm 2.92}$ \\
            Logsum EPKL & $42.40_{\pm 2.92}$ \\
            Mean RMI & $42.38_{\pm 2.92}$ \\
            Logsum EoE & $42.37_{\pm 2.92}$ \\
            Mean ExE & $42.25_{\pm 2.90}$ \\
            Logsum ExE & $42.22_{\pm 2.90}$ \\
            Logsum RMI & $42.10_{\pm 2.89}$ \\
            Logsum NC & $41.64_{\pm 2.86}$ \\
            Logsum MI & $41.26_{\pm 2.79}$ \\
            Random & $41.13_{\pm 2.78}$ \\
            \bottomrule
        \end{tabular}}
    \end{subtable}
    \caption{
    Average across the test set areas under DSC-RC (a) and Lesion F1-RC (b) measuring the correspondence between uncertainty measures and model errors in segmentation and lesion detection, respectively. Standard errors are computed using bootstrapping with the sample size of 50 patients for $10^{4}$ repetitions.
    }
    \label{tab:aac}
\end{table}

\clearpage

\section{Compliance with ethical standards}
\label{sec:ethics}
This research study was conducted retrospectively using human subject data made available in open access by the Shifts project~\cite{shifts20}. Ethical approval was not required as confirmed by the license attached with the open data.
\section{Acknowledgments}
\label{sec:acknowledgments}
This work was supported by the Hasler Foundation Responsible AI programme (MSxplain) and the EU Horizon 2020 project AI4Media (grant 951911). We acknowledge access to the facilities and expertise of the CIBM Center for Biomedical Imaging, a Swiss research center of excellence founded and supported by Lausanne University Hospital (CHUV), University of Lausanne (UNIL), École polytechnique fédérale de Lausanne (EPFL), University of Geneva (UNIGE) and Geneva University Hospitals (HUG).

\bibliographystyle{IEEEbib}
\bibliography{main}

\end{document}